\documentclass[aps,prd,twocolumn,groupedaddress,showpacs,amsmath,amssymb]{revtex4}
    \newcommand{\be}[1]{\begin{equation}\label{#1}}
    \newcommand{\ep}[1]{\epsilon_{#1}}
    \newcommand{\de}[1]{\delta_{#1}}

    \newcommand{\pa}[1]{\left(#1\right)}
    \newcommand{\paq}[1]{\left[#1\right]}

    \newcommand{\ab}[1]{\left|#1\right|}
    
    \newcommand{\PR}{\mathcal{P}_{\mathcal{R}}}
    \def\ee{\end{equation}}
    \def\ba{\begin{eqnarray}}
    \def\ea{\end{eqnarray}}

\usepackage{graphicx,latexsym}
\usepackage{graphicx,epsf}
\usepackage{color}
\usepackage{epsfig}
\begin{document}
\title{Quantum Back-Reaction in Scale Invariant Induced Gravity Inflation}
\author{A. 
Tronconi$\,^{1}$ and G. Venturi$\,^{1,2}$}

\affiliation{$^{1}$ Dipartimento di Fisica, Universit\`a degli Studi di
Bologna, via Irnerio, 46 -- I-40126 Bologna -- Italy}
\affiliation{$^{2}$ INFN, Sezione di Bologna,
Via Irnerio 46, I-40126 Bologna, Italy}
\begin{abstract}
A quartic, self-interacting potential in the induced gravity framework is known to have a pure de Sitter attractor for the homogeneous mode. In order to obtain non-zero slow roll parameters we therefore study the quantum back-reaction of the scalar and the tensor perturbations on such a homogeneous dynamics. The results are then compared with inflationary observables in order to constrain the parameters of the model.
\end{abstract}
\pacs{98.80Cq}
\maketitle
\section{Introduction}
The paradigm of inflation connects primordial quantum fluctuations to the inhomogeneities observed in the large scale structures and in the Cosmic Microwave background (CMB). Indeed the variation of the cosmological background due to the accelerated expansion of the universe and the resulting non-adiabaticity excites the vacuum and leads to the production of matter quanta seeding the primordial inhomogeneities. Similarly gravitational waves can be produced in this era. The subsequent gravitational collapse of the matter thus created leads to the observed large scale structures. The ever increasing precision of satellite experiments in measuring the power spectrum associated with the distributions of inhomogeneities will lead to the understanding of the microscopic quantum mechanical mechanism which drives inflation and leads to the primordial structures. Thus just as one has a standard model for elementary particle physics one searches for a ``standard model'' for inflation and the distribution of matter in the universe.\\
Many models for inflation have been proposed (see \cite{Lyth:1998xn} for a review) and in such models matter is generally described by a scalar field (inflaton) coupled to gravity in diverse ways. Many years ago a simple scalar field model for the generation of Newton's constant through the spontaneous breaking of scale invariance in a curved space was presented \cite{CV} and its simple cosmological consequences studied. In particular the model, besides generating the gravitational constant, was found to have a de Sitter attractor in the future \cite{FTV}.\\
In subsequent papers \cite{CFTV,Cerioni} we studied in detail the inflationary dynamics and reheating for induced gravity models with different symmetry-breaking potentials.
In particular the potentials we examined were associated with the presence of a condensate (Landau-Ginzburg) or quantum effects (Coleman-Weinberg). In particular for the latter case we used an effective potential inspired by the result obtained in flat-space.\\ 
The scope of this manuscript is to examine the predictions obtained from the effective Lagrangian which includes the quantum correction evaluated on a curved (de Sitter) space time. In Section II induced gravity (IG) is briefly reviewed. Section III is dedicated to the scalar cosmological perturbations, their regularization and renormalization for both the Bunch-Davies and Allen-Folacci vacuum. In section IV the gravitational results are similarly studied. In section V the total back-reaction is studied and compared to observations and in section VI conclusions are drawn.
\section{IG inflation}
We consider the IG model \cite{sakharov} described by the action
\begin{equation}
S = \int d^4 x \sqrt{-g}\left[-\frac{g^{\mu\nu}}{2}\partial_{\mu}
\sigma\partial_{\nu}\sigma+{\gamma\over 2}\sigma^2 R-\frac{\lambda}{4!}\sigma^{4} \right]
\label{action}
\end{equation}
where $\sigma$ now is the inflaton scalar field, $\gamma$ and $\lambda$ are dimensionless, positive definite parameters representing the non-minimal coupling between the scalar field and gravity and the scalar field self-coupling respectively. The Einstein-Hilbert term for gravity is replaced by an effective mass-like term for the scalar field $\sigma$. When the inflaton energy density dominates it drives the dynamics of the space-time. Further an expectation value of the scalar field will play the role of an effective Planck mass. On restricting our analysis to the homogeneous mode of the scalar field and assuming a spatially flat Robertson-Walker background 
\be{metric}
d s^2 = g_{\mu \nu} dx^{\mu} dx^{\nu}=-dt^2+a^2(t) d\vec{x}^2 \,,
\ee
the variation of the above Lagrangian leads to the following set of 
independent equations
\begin{eqnarray}
\!\!\!\!&3\gamma H^2& \pa{1+2\de{1}-\frac{1}{6\gamma}\de{1}^{2}}= \frac{\lambda}{24}\sigma^{2}\label{homeq1}\\
&3\gamma H^{2}&\paq{1+\frac{\de{1}^{2}}{6\gamma}+\frac{2}{3}\de{1}\pa{2+3\de{1}-\ep{1}-\frac{\ep{1}}{\de{1}}}}\nonumber\\
&&=\frac{\lambda}{24}\sigma^{2}\label{homeq2}
\end{eqnarray}
in terms of the slow-roll (SR) parameters $\de{n}$ and $\ep{n}$ where $H\equiv \dot a/a$, $\sigma=\sigma(t)$ and the dot is the derivative w.r.t. the cosmic time $t$. The SR parameters are defined recursively for $n \ge 0$ by $d\ln |\delta_n|/dN \equiv \delta_{n+1}$, $\delta_0 \equiv \sigma/\sigma(t_{i})$ and $d\ln |\epsilon_n|/dN \equiv \epsilon_{n+1}$, $\epsilon_0 \equiv H(t_{i})/H$ with $n \ge 0$, where
$t_{i}$ is some initial time and $N\equiv\ln \frac{a}{a(t_{i})}$
is the number of e-folds.
The homogeneous dynamics of the above set of equations has been studied in detail in our previous papers \cite{FTV,Cerioni} and is known to have de Sitter attractors with
\be{hsol}
H^{2}=\lambda \frac{\sigma^{2}}{72\gamma},\quad\de{n}=0,\quad\ep{n}=0.
\ee
On the other hand such a background dynamics (\ref{hsol}) is unable to reproduce correctly the small deviations from scale invariance observed in the spectrum of perturbations generated during inflation.\\
Our goal in this paper is to evaluate the deviations from (\ref{hsol}) due to the quantum back-reaction on the homogeneous dynamics of the scalar and tensor perturbations produced by the accelerated expansion. Such an approach is more suitable in order to take into account the quantum corrections to the homogeneous classical dynamics of the inflaton. Making use of an ``ad hoc'' flat-space Coleman-Weinberg effective potential in order to reproduce the effect of quantum fluctuations is inappropriate for the early universe when the curvature of space-time is non negligible and the Universe is filled with nothing more than inflaton energy.
The modification of the SR parameters will affect the spectrum of scalar and tensor perturbations produced during inflation and will be compared with observations.\\
Let us end this section by emphasizing that we shall evaluate the quantum corrections in a de Sitter phase, that is on the attractor of the background dynamics. Strictly speaking one actually approaches the attractor and the SR parameters are non-zero. Indeed if we perturb the homogeneous solution as \cite{CV}
\be{pertCV}
\sigma_{0}\pa{t}=\sigma_{0}\pa{1+\xi(t)}\,,\quad a=s_{0}(t)+s(t)
\ee
our equations (\ref{homeq1},\ref{homeq2}) can be expanded for $|\xi|\ll1$ and $|s|\ll |s_{0}|$ and take the form
\be{expVC}
\frac{\dot s-H s}{s_{0}}=-\dot\xi+ H\xi\,,\quad\ddot \xi+3 H\dot \xi=0.
\ee
On requiring that $s_{0}(t)$ (pure de Sitter) be the asymptotic solution and that $\xi(t)$ be zero for $t\rightarrow\infty$ one has
\be{sCV1}
\xi(t)=-\frac{\dot\xi(0)}{3 H}{\rm e}^{-3 Ht}
\ee
and
\be{sCV2}
a(t)=s_{0}(0){\rm e}^{H t}+\frac{4}{9 H}s_{0}(0)\dot\xi(0){\rm e}^{-2 Ht}.
\ee
Thus as a consequence our SR parameters are non-zero and in particular
\be{de1CV}
\de{1}=\frac{\dot\xi(0)}{H}{\rm e}^{-3 Ht}
\ee
and
\be{ep1CV}
\ep{1}=-4\frac{\dot \xi(0)}{H}{\rm e}^{-3Ht}.
\ee
The smallness of the SR parameters and of the corrections to pure de Sitter in (\ref{pertCV}) require $\|\dot\xi(0)\|\ll H$. We see that in the absence of matter (see \cite{FTV}), even at the end of inflation ($s\ll s_{0}$), we are still not in a pure de Sitter stage. In the subsequent sections we shall work to lowest order for the homogeneous part, that is ignore the deviation from pure de Sitter which we expect to at most be of the same order of magnitude as the corrections due to the quantum back-reaction.

\section{Scalar Perturbations}
The dynamics of the scalar perturbations in an expanding Universe has been studied in detail in many papers \cite{Mukhanov:1990me,Hwang95,Noh01}. The scalar perturbations of the metric tensor contain two unphysical degrees of freedom which can be eliminated by a suitable gauge transformation (coordinate transformation). Finally the physical perturbations can be conveniently expressed in terms of gauge invariant combinations of the remaining degrees of freedom. This prescription is necessary both for calculating the spectra of the cosmological perturbations and for studying the problem of the back-reaction of the perturbations on the homogeneous dynamics. The calculation of the quantum back-reaction on curved space is a delicate task. The results are know to depend on the gauge choice and a universally accepted prescription does not appear to exists (see for instance \cite{Gasperini:2009mu,Li:2007ci,Finelli:2006wk} for recent discussions of this problem). In this paper we do not address the issue of gauge dependence of the quantum back-reaction and shall perform the calculations in a specific gauge.\\
In particular, in the IG framework, the uniform curvature gauge (UCG) appears to be a simple choice to eliminate the unphysical quantities and fix the gauge \cite{Hwang95,Noh01}.\\
In general, the scalar perturbations of the metric for the homogeneous and isotropic FRW background (\ref{metric}) can be written as
\be{pertmet}
ds^2=-\pa{1+2\alpha}dt^2-\chi_{,i} \,dt\;dx^i+a^2 \delta_{ij}\pa{1+2\varphi}dx^i\;dx^j
\ee
where the spatial gauge has already been fixed and the fields $\alpha(\vec x,t)$, $\chi(\vec x,t)$, $\varphi(\vec x,t)$ are temporal gauge dependent. Furthermore, for our model, one has to consider the inflaton fluctuation $\delta \sigma(\vec x,t)$ defined as $\sigma(\vec x,t)=\sigma_0(t)+\delta\sigma(\vec x,t)$ which is temporal gauge dependent as well.
The UCG consists in setting $\varphi=0$ by using the temporal gauge freedom. One is then left with 3 fields for the perturbations, $\alpha$, $\chi$ and $\delta\sigma$, which are coupled at the linear order through the Einstein equations and the Klein-Gordon (KG) equation for the inflaton field. In terms of these functions the full action (\ref{action}) can be written as the sum of two contributions: the homogeneous part, $S_h$, and the scalar perturbations action, $S_{sp}$, which is quadratic in $\delta\sigma$, $\alpha$, $\chi$ and their derivatives:
\be{fullaction}
S\simeq S_h+S_{sp}
\ee
where
\be{homaction}
S_{h}=\int d^4x\; a^3\paq{\frac{\dot\sigma_0^2}{2}+3\gamma\sigma_0^2 \pa{\frac{\dot a^2+a\ddot a}{a^2}}-\frac{\lambda}{4!}\sigma_0^4}
\ee
and $S_{sp}$ has a quite involved structure.\\
The variation of $S$ w.r.t. the scalar perturbations and the subsequent use of the homogeneous solution (\ref{hsol}) leads to:
\be{sccon1}
\alpha+\frac{\delta\sigma}{\sigma_{0}}-\frac{\delta\dot\sigma}{H\sigma_{0}}=0,
\ee
\begin{eqnarray}
&&\frac{\lambda\,\sigma_{0}^{2}}{72\gamma H^{2}}\pa{\alpha+\frac{\delta\sigma}{\sigma_{0}}}
+\frac{\delta^{ij}\pa{\delta\sigma_{,ij}\sigma_{0}^{-1}-aH\chi_{,ij}}}{3a^{2}H^{2}}\nonumber\\
&&-\frac{\delta\dot\sigma}{H\sigma_{0}}=0\label{sccon2}
\end{eqnarray}
and
\begin{eqnarray}
&&\frac{\delta\ddot\sigma}{\sigma_{0}}+\frac{2\gamma\delta^{ij}}{a^{2}}\pa{\alpha-\frac{\delta\sigma}{2\gamma\sigma_{0}}-3aH\chi-a\dot\chi}_{,ij}\nonumber\\
&&+\frac{\lambda\,\sigma_{0}^{2}}{3}\pa{\alpha+\frac{\delta\sigma}{\sigma_{0}}}+H\pa{6\gamma\dot\alpha+3\frac{\delta\dot\sigma}{\sigma_{0}}}=0\label{sccon3}
\end{eqnarray}
which represents the linear constraints which relate $\alpha$, $\chi$ and $\delta\sigma$ and the comma denotes the partial derivative $f_{,i}\equiv\partial f/\partial x^i$, .
The variation of $S$ w.r.t. $n(t)$, $a(t)$ and $\sigma_{0}(t)$ gives the Einstein equations (\ref{homeq1},\ref{homeq2}) and the KG equation toghether with the quadratic corrections coming from the scalar perturbations.\\
On using (\ref{sccon1},\ref{sccon2},\ref{sccon3}) it is possible to reduce the scalar perturbations in $S_{sp}$ to a single, gauge invariant, degree of freedom, $\delta\sigma_\varphi\equiv\delta\sigma+\frac{\dot\sigma_0}{H}\varphi$, which is the inflaton fluctuation $\delta\sigma$ itself in the UCG. The action $S_{sp}$ in terms of $\delta\sigma$ has been calculated for a generic potential in the IG framework \cite{Hwang:1996bc} and has the following general form
\begin{eqnarray}
S_{\delta\sigma}=\frac{1}{2}\int d^4x\;a^3 Z\left\{\delta\dot\sigma^2-\frac{1}{a^2}\delta^{ij}\delta\sigma_{,i} \delta\sigma_{,j}\right.\nonumber\\
\left.+\frac{H}{a^3\dot\sigma_0}Z^{-1}\frac{d}{dt}\paq{a^3Z\frac{d}{dt}\pa{\frac{\dot \sigma_0}{H}}}\delta\sigma^2\right\}=0\label{pertaction}
\end{eqnarray}
where 
\be{Z}
Z={\pa{1+6\gamma}\over\pa{1+\de{1}}^2}.
\ee
The form of the equations for the homogeneous degrees of freedom with the perturbations is much more involved. Some simplification occurs when one evaluates them perturbatively by using (\ref{hsol}) in the second order contributions. The equations one finally obtains contain contributions up to the fourth order in the space-time derivatives, and a linear combination of them gives the following
\begin{widetext}
\begin{eqnarray}
&&\de{1}\pa{1+6\gamma}\pa{3+2\de{1}+\de{2}-\ep{1}}=-\frac{1+6\gamma}{\sigma_{0}^{2}H^{2}}\paq{5\delta\dot\sigma^{2}+\delta^{ij}\frac{\delta\sigma_{,i}\,\delta\sigma_{,j}}{a^{2}}+2\delta^{ij}\frac{\delta\sigma\,\delta\sigma_{,ij}}{a^{2}}+2\delta^{ij}\frac{\delta\sigma_{,i}\,\delta\dot\sigma_{,j}}{a^{2}H}-2\delta^{ij}\frac{\delta\dot\sigma\,\delta\sigma_{,ij}}{a^{2}H}},\label{eqp1}\\
&&2\pa{\de{1}+\ep{1}}-\de{1}\paq{\frac{4\gamma+1}{\gamma}\de{1}+2\pa{\de{2}-\ep{1}}}=-\frac{1}{\sigma_{0}^{2}H^{2}}\paq{\frac{6\gamma-1}{\gamma}\delta\dot\sigma^{2}-\frac{1+4\gamma}{3\gamma}\delta^{ij}\frac{\delta\sigma_{,i}\,\delta\sigma_{,j}}{a^{2}}+8\delta^{ij}\frac{\delta\sigma\,\delta\sigma_{,ij}}{a^{2}}\right.\nonumber\\
&&\left.-12H\frac{d\delta\sigma^{2}}{dt}+\frac{4}{3}\delta^{ij}\frac{\delta\sigma_{,i}\,\delta\dot\sigma_{,j}}{a^{2}H}-\frac{10}{3}\delta^{ij}\frac{\delta\dot\sigma\,\delta\sigma_{,ij}}{a^{2}H}-\frac{2}{3}\delta^{ij}\pa{\frac{\delta\dot\sigma_{,i}\,\delta\dot\sigma_{,j}}{a^{2}H^{2}}-\frac{\delta\ddot\sigma\,\delta\sigma_{,ij}}{a^{2}H^{2}}-\delta^{kl}\frac{\delta\sigma_{,ij}\delta\sigma_{,kl}-\delta\sigma_{,ik}\delta\sigma_{,jl}}{a^{4}H^{2}}}}.\label{eqp2}
\end{eqnarray}
\end{widetext}
In order to evaluate the back-reaction of the scalar perturbations on the homogeneous dynamics, $\delta\sigma$ on r.h.s of Eqs. (\ref{eqp1},\ref{eqp2}) has to be quantized and evaluated for a suitable vacuum state \cite{Finelli:2001bn,Finelli:2003bp}.\\
The equation governing the dynamics of the inflaton fluctuation follows from (\ref{pertaction}) and, mode by mode, is given by
\ba
&&\frac{d^{2}\delta\tilde\sigma(\vec k,t)}{dt^{2}}+\left( 3 H + \frac{\dot Z}{Z} \right)
\frac{d\delta\tilde\sigma(\vec k,t)}{dt} \nonumber\\
&&+\left[ \frac{k^2}{a^2} - \frac{1}{a^3 Z \sigma_{0} \delta_1}
\left( a^3 Z \left( \sigma_{0} \delta_1 \right)^\cdot \right)^\cdot
\right] \delta\tilde\sigma(\vec k,t) = 0\label{scperteq}
\ea
where $\delta\tilde\sigma(\vec k,t)$ is the Fourier amplitude defined by $\delta\sigma(\vec x,t)=\pa{2\pi}^{-3/2}\int d^3k\;\exp\paq{i\vec k\cdot \vec x}\delta\tilde\sigma(\vec k,t)$. Such an amplitude can be quantized and expressed through the creation-annihilation operators as $\delta\tilde\sigma(\vec k,t)\equiv \delta\sigma_{k}(t)\, \hat a_{\vec k}+\delta\sigma_{k}^{*}(t)\,\hat a^{\dagger}_{-\vec k}$. When evaluated for a vacuum state $|0\rangle$ (defined as usual by $\hat a_{\vec k}|0\rangle=0$) the r.h.s. of Eqs. (\ref{eqp1},\ref{eqp2}) can be re-written in terms of $\delta\sigma_{k}$ in the following form
\begin{eqnarray}
&&-\frac{1+6\gamma}{\sigma_{0}^{2}H^{2}}\frac{1}{2\pi^{2}}\int dk\;k^{2}\pa{5\ab{\delta\dot\sigma_{k}}^{2}-\frac{k^{2}}{a^{2}}\ab{\delta\sigma_{k}}^{2}\right.\nonumber\\
&&\left.+2\frac{k^{2}}{a^{2}H}\frac{d\ab{\delta\sigma_{k}}^{2}}{dt}}\label{rhsp1}
\end{eqnarray}
and
\begin{widetext}
\be{rhsp2}
\frac{1}{\sigma_{0}^{2}H^{2}}\frac{1}{2\pi^{2}}\int dk\;k^{2}\paq{\frac{1-6\gamma}{\gamma}\ab{\delta\dot\sigma_{k}}^{2}+\frac{1+28\gamma}{3\gamma}\frac{k^{2}}{a^{2}}\ab{\delta\sigma_{k}}^{2}+12H\frac{d\ab{\delta\sigma_{k}}^{2}}{dt}-\frac{7}{3}\frac{k^{2}}{a^{2}H}\frac{d\ab{\delta\sigma_{k}}^{2}}{dt}+\frac{1}{3}\frac{k^{2}}{a^{2}H^{2}}\frac{d^{2}\ab{\delta\sigma_{k}}^{2}}{dt^{2}}}
\ee
\end{widetext}
respectively.\\
The above integrals are divergent in the ultraviolet limit and must be regularized and renormalized so as to obtain finite results. In this paper we shall adopt dimensional regularization and adiabatic subtraction to cure these divergences.\\ 
The equation governing the quantum dynamics of the scalar fluctuation (\ref{scperteq}) can be cast into the form of a massless test field on a quasi-de Sitter background or, equivalently, of a massive test field on a de Sitter background with the ``mass'' being a function of the SR parameters. In our perturbative approach we shall evaluate (\ref{scperteq}) on the zeroth order background solution (\ref{hsol}) and the ``effective mass'' of the scalar perturbations is then set to zero. The Allen-Folacci (AF) vacuum choice \cite{Allen:1987tz} is known to be the correct prescription for the treatment of the zero mode of a massless scalar field on an exact de Sitter space. With such a vacuum the expressions (\ref{rhsp1},\ref{rhsp2}) are free of infrared divergences.\\
The Bunch-Davies (BD) vacuum choice \cite{Bunch:1978yq} for a massive scalar field leads to different results to the AF prescription when the massless limit is taken at the end of the calculations \cite{Finelli:2001bn} and is not the correct choice when dealing with massless test fields. Let us stress, however, that, when the effect of the back-reaction is considered, the SR parameters turn to be non zero and the scalar perturbations do have an ``effective mass''. The BD vacuum choice appears to be therefore the correct choice for quantizing the scalar dynamics. At the end of the calculations we shall then set the mass to zero because it is associated with higher order contributions to (\ref{eqp1},\ref{eqp2}).\\
In the following we perform the calculations for both vacuum prescriptions in order to compare them bearing in mind that the massless limit of the BD vacuum appears to be the correct choice.
For the BD choice, one introduces a mass regulator in the action (\ref{pertaction})
\be{massaction}
\Delta S_{\delta\sigma}=-\frac{m^{2}}{2}\int d^4x\;a^3 Z\;\delta\sigma^2
\ee
thus eliminating the infrared singularity and leading to a modified KG equation which can be simply obtained from (\ref{scperteq}) by replacing $k^{2}/a^{2}\rightarrow m^{2}+k^{2}/a^{2}$. The introduction of the mass regulator (\ref{massaction}) modifies the constraint (\ref{sccon3}) but leaves (\ref{sccon1},\ref{sccon2}) unchanged. Furthermore it affects the Friedmann, the acceleration and the KG equations for the homogeneous degrees of freedom to second order in $\delta\sigma_{k}$. Consequently contributions proportional to $m^{2}$ appear on the r.h.s. of (\ref{rhsp1}):
\be{rhsp1reg}
-\frac{1+6\gamma}{\sigma_{0}^{2}H^{2}}\frac{1}{2\pi^{2}}\int dk\;k^{2}\,m^{2}\paq{\frac{3}{2H}\frac{d\ab{\delta\sigma_{k}}^{2}}{dt}-\ab{\delta\sigma_{k}}^{2}}
\ee
and of (\ref{rhsp2}):
\be{rhsp2reg}
-\frac{1}{\sigma_{0}^{2}H^{2}}\frac{1}{2\pi^{2}}\int dk\;k^{2}\,m^{2}\paq{\frac{1}{H}\frac{d\ab{\delta\sigma_{k}}^{2}}{dt}-6\ab{\delta\sigma_{k}}^{2}}
\ee
respectively.
At the end of the calculations one finally eliminates the dependence on the regulator by taking the $m\rightarrow 0$ limit.
\subsection{Regularization and Renormalization}
The dynamics of scalar perturbations on a FRW background is given by Eq. (\ref{scperteq}). Our approximation scheme consists on expanding the background dynamics in series as $\sigma_{0}(t)=\sigma_{0}^{(0)}+\sigma_{0}^{(1)}(t)+\dots$ and $H(t)=H^{(0)}+H^{(1)}(t)+\dots$ where $\sigma_{0}^{(0)}$ and $H^{(0)}$ satisfy the unperturbed equations (\ref{hsol}) and are thus constant, while $\sigma_{0}^{(1)}(t)$ and $H^{(1)}(t)$ depend on the perturbations. In this scheme the slow-roll parameters on the l.h.s. (\ref{eqp1},\ref{eqp2}) are next to leading order quantities and will be determined by the perturbations. Let us note that $\sigma_{0}^{(1)}(t)$ and $H^{(1)}(t)$ will also obtain ``off-attractor'' contributions such as (\ref{de1CV},\ref{ep1CV}) which will rapidly decrease and can be safely neglected for our proposes.\\
Within this perturbative approach Eq. (\ref{scperteq}) (with the mass regulator (\ref{massaction})) greatly simplifies and becomes
\be{scpert}
\delta\ddot\sigma_{k}+3H\delta\dot\sigma_{k}+\pa{\frac{k^{2}}{a^{2}}+m^{2}}\delta\sigma_{k}=0
\ee
where $a(t)=a_{0}\exp H^{(0)}t$.\\
The perturbations dynamics can be quantized by introducing the conformal time $\eta$ ($dt=a\;d\eta$) and the rescaled scalar perturbation $\psi_{\vec k}=\sqrt{1+6\gamma}\,a\,\delta\tilde\sigma(\vec k,t)$  \cite{Birrell:1982ix}. In terms of $\psi_{\vec k}$ the action (\ref{pertaction}) is canonically normalized and takes the following form
\be{psiscaction}
S_{\psi}=\frac{1}{2}\int d^{3}k\,d\eta\paq{(\psi_{\vec k}')^{2}-\pa{k^{2}+a^{2}m^{2}-2\mathcal{H}^{2}}\psi_{\vec k}^{2}}
\ee
where the prime denotes the derivative w.r.t. the conformal time and $\mathcal{H}\equiv a'/a=a H^{(0)}$. The variation of (\ref{psiscaction}) leads to the KG equation
\be{eqpsisc}
\psi''_{\vec k}+\paq{k^{2}-\frac{2}{\eta^{2}}\pa{1-\frac{m^{2}}{2H^{2}}}}\psi_{\vec k}=0
\ee 
where we shall take $H=H^{(0)}={\rm const}$. In terms of the creation and anihilation operators the scalar field $\psi_{\vec k}$ can be expanded as
\be{cran}
\psi_{\vec k}=f_{k}(\eta)\,\hat a_{\hat k}+f^{*}_{k}(\eta)\,\hat a^{\dagger}_{-\vec k}
\ee
where $f_{k}(\eta)=\sqrt{1+6\gamma}\,a\,\delta\sigma_{k}$ is normalized so that
\be{normf}
f_{k}\pa{f^{*}_{k}}'-\pa{f_{k}}'f^{*}_{k}=i.
\ee
and satisfies (\ref{eqpsisc}).
The normalization condition determines just one of the two integration constant in $f_{k}$. A second condition is needed in order to fix uniquely the vacuum state $a_{\vec k}|0\rangle=0$. The BD vacuum corresponds to the condition $f_{k}(\eta)\rightarrow \exp\paq{ik\eta}$ when $k\eta\rightarrow -\infty$ and leads to
\be{BDvac}
f_{k}=\frac{\sqrt{-\pi \eta}}{2}H_{\nu}^{(1)}(-k\eta)
\ee
where $H_{\nu}^{(1)}(z)$ is the Hankel function and $\nu=\sqrt{\frac{9}{4}-\frac{m^{2}}{H^{2}}}$.\\
The integrals in (\ref{rhsp1},\ref{rhsp2}) plus the corrections (\ref{rhsp1reg},\ref{rhsp2reg}), evaluated for the BD vacuum, are divergent in the ultraviolet and one must choose a renormalization scheme to obtain finite results. In particular we choose dimensional regularization and fourth order adiabatic subtraction \cite{Finelli:2001bn}.\\
\subsection{BD Results}
One then obtains the following renormalized expressions for the BD vacuum choice in the massless limit:
\be{m2s2BD}
\frac{1}{2\pi^{2}}\int dk\,k^{2}\pa{m^{2}\ab{\delta\sigma_{k}}^{2}}\rightarrow\frac{61H^{4}}{240\pa{1+6\gamma}\pi^{2}},
\ee
\begin{eqnarray}
&&\frac{1}{2\pi^{2}}\int dk\,k^{2}\paq{\ab{\delta\dot\sigma_{k}}^{2}+\pa{\frac{k^{2}}{a^{2}}+m^{2}}\ab{\delta\sigma_{k}}^{2}}\nonumber\\
&&\rightarrow\frac{61H^{4}}{480\pa{1+6\gamma}\pi^{2}},\label{2rhoBD}
\end{eqnarray}
\be{k2a2s2BD}
\frac{1}{2\pi^{2}}\int dk\,k^{2}\pa{\frac{k^{2}}{a^{2}}\ab{\delta\sigma_{k}}^{2}}\rightarrow-\frac{61H^{4}}{320\pa{1+6\gamma}\pi^{2}}.
\ee
Let us note that the above results are as expected apart from the factor $1+6\gamma$ in the denominator which follows from the normalization of $\psi_{k}$ \cite{Finelli:2001bn}.\\
On using the fact that, owing to the perturbative approach employed, the r.h.s. of (\ref{m2s2BD},\ref{2rhoBD},\ref{k2a2s2BD}) are constant and taking the derivative of the result (\ref{m2s2BD}) w.r.t. the cosmic time $t$ one obtains
\be{sdsBD}
\frac{1}{2\pi^{2}}\int dk\,k^{2}\pa{\frac{d\,\ab{\delta\sigma_{k}}^{2}}{dt}}\rightarrow0.
\ee
Further, on also deriving (\ref{k2a2s2BD}), one finds that the following relations hold for the integrals and for their renormalized results as well:
\begin{widetext}
\be{reldt}
0=\frac{d}{dt}\paq{\int dk\,k^{2}\pa{\frac{k^{2}}{a^{2}}\ab{\delta\sigma_{k}}^{2}}}=-2H\int dk\,k^{2}\pa{\frac{k^{2}}{a^{2}}\ab{\delta\sigma_{k}}^{2}}+\int dk\,k^{2}\pa{\frac{k^{2}}{a^{2}}\frac{d\ab{\delta\sigma_{k}}^{2}}{dt}},
\ee
\be{relddt}
0=\frac{d^{2}}{dt^{2}}\paq{\int dk\,k^{2}\pa{\frac{k^{2}}{a^{2}}\ab{\delta\sigma_{k}}^{2}}}=-4H^{2}\int dk\,k^{2}\pa{\frac{k^{2}}{a^{2}}\ab{\delta\sigma_{k}}^{2}}-4H\int dk\,k^{2}\pa{\frac{k^{2}}{a^{2}}\frac{d\ab{\delta\sigma_{k}}^{2}}{dt}}+\int dk\,k^{2}\pa{\frac{k^{2}}{a^{2}}\frac{d^{2}\ab{\delta\sigma_{k}}^{2}}{dt^{2}}}
\ee
\end{widetext}
leading to
\be{k2a2HsdsBD}
\frac{1}{2\pi^{2}}\int dk\,k^{2}\pa{\frac{k^{2}}{a^{2}H}\frac{d\,\ab{\delta\sigma_{k}}^{2}}{dt}}\rightarrow-\frac{61H^{4}}{160\pa{1+6\gamma}\pi^{2}}
\ee
and
\be{k2a2HddssBD}
\frac{1}{2\pi^{2}}\int dk\,k^{2}\pa{\frac{k^{2}}{a^{2}H^{2}}\frac{d^{2}\,\ab{\delta\sigma_{k}}^{2}}{dt^{2}}}\rightarrow-\frac{183H^{4}}{80\pa{1+6\gamma}\pi^{2}}.
\ee
We note that the results given by (\ref{k2a2HsdsBD}, \ref{k2a2HddssBD}) are the consequence of the fact that the expression (\ref{k2a2s2BD}) is constant.
\subsection{AF Results}
A quite different approach is followed in order to renormalize the divergent integrals obtained for the AF vacuum choice. In such a case the mass regulator is not necessary and one can eliminate the infrared and ultraviolet divergences in (\ref{rhsp1},\ref{rhsp2}) by subtracting mode by mode from the singular contributions the corresponding terms in the adiabatic series before performing the integrals \cite{Finelli:2001bn}.\\
In the massless case, the KG equation for the perturbations (\ref{scpert}) has two independent solutions which can be written as
\be{AFvac}
f_{k}=\frac{1}{\sqrt{2k}}\pa{1-\frac{i}{k\eta}}{\rm e}^{-ik\eta}
\ee
and its complex conjugate. Mode by mode subtraction of the divergent quantities in (\ref{rhsp1},\ref{rhsp2}) leads to the following renormalized expressions
\be{2rhoAF}
\frac{1}{2\pi^{2}}\int dk\,k^{2}\pa{\ab{\delta\dot\sigma_{k}}^{2}+\frac{k^{2}}{a^{2}}\ab{\delta\sigma_{k}}^{2}}\rightarrow-\frac{119H^{4}}{480\pa{1+6\gamma}\pi^{2}},
\ee
\be{k2a2s2AF}
\frac{1}{2\pi^{2}}\int dk\,k^{2}\pa{\frac{k^{2}}{a^{2}}\ab{\delta\sigma_{k}}^{2}}\rightarrow-\frac{119H^{4}}{320\pa{1+6\gamma}\pi^{2}},
\ee
The relations (\ref{reldt},\ref{relddt}) still hold and one finds  
\be{k2a2HsdsAF}
\frac{1}{2\pi^{2}}\int dk\,k^{2}\pa{\frac{k^{2}}{a^{2}H}\frac{d\,\ab{\delta\sigma_{k}}^{2}}{dt}}\rightarrow-\frac{119H^{4}}{160\pa{1+6\gamma}\pi^{2}}
\ee
and
\be{k2a2HddssAF}
\frac{1}{2\pi^{2}}\int dk\,k^{2}\pa{\frac{k^{2}}{a^{2}H^{2}}\frac{d^{2}\,\ab{\delta\sigma_{k}}^{2}}{dt^{2}}}\rightarrow-\frac{357H^{4}}{80\pa{1+6\gamma}\pi^{2}}.
\ee
Let us note that (\ref{2rhoAF},\ref{k2a2s2AF}) are the usual results for a massless scalar field on a de Sitter background \cite{Finelli:2001bn}.
\section{Gravitational Waves}
Gravitational waves can be produced during inflation and their contribution must be accounted for when calculating the quantum back-reaction on the homogeneous dynamics \cite{Finelli:2004bm}. Gravitational waves (i.e. tensor perturbations of the FRW metric) are gauge invariant perturbations described by the traceless, transverse tensor $h_{ij}$ defined by
\be{tenspertmet}
ds^2=-dt^2+a^2 \pa{\delta_{ij}+h_{ij}}dx^i\;dx^j
\ee
At the linear order in their amplitude the dynamics of the tensor perturbations decouples from that of the scalar perturbations and can be described by the action
\be{gwaction}
S_{gw}=\sum_{\alpha=+,\times}\frac{\gamma}{4}\int d^{4}x\;a^{3}\,\sigma_{0}^{2}\paq{\dot h_{\alpha}^2-\frac{1}{a^2}\delta^{ij}h_{\alpha,i} h_{\alpha,j}-\mu^2_Th_{\alpha}^2}
\ee
where $\alpha$ describes the two independent degrees of freedom in $h_{ij}$ related to the possible polarizations of the graviton and
\be{muT}
\mu^2_T\equiv2\frac{\dot a^2}{a^2}+4\frac{\ddot a}{a}+8\frac{\dot a}{a}\frac{\dot\sigma_0}{\sigma_0}+\frac{4\gamma+1}{\gamma}\frac{\dot \sigma_0^2}{\sigma_0^2}+4\frac{\ddot \sigma_0}{\sigma_0}+m^{2}
\ee
where $m^{2}$ is the mass regulator added to cure the infrared singularity for the BD vacuum choice. Let us note that tensor perturbations (just as the scalar perturbations) again have an ``effective mass'' when the SR parameters are different from zero. For a comparison the quantization will again be performed both for the BD and AF procedures.\\
The full action (\ref{fullaction}) with scalar and tensor contributions is
\be{fullactionGW}
S=S_{h}+S_{sp}+S_{gw}
\ee
and its variation leads to Eqs. (\ref{eqp1},\ref{eqp2}) plus the tensor contributions which, on r.h.s., take the following form
\be{eqp1gw}
\frac{\gamma}{2}\frac{m^{2}}{H^{2}}\sum_{\alpha=+,\times}h_\alpha^2,
\ee
\begin{eqnarray}
&&\frac{1}{6H^{2}}\sum_{\alpha=+,\times}\pa{-\dot h_{\alpha}^2+\delta^{ij}\frac{h_{\alpha,i}h_{\alpha,j}}{a^2}-4\,\delta^{ij}\frac{h_{\alpha}h_{\alpha,ij}}{a^2}\right.\nonumber\\
&&\left.+6H\frac{d\,h_\alpha^2}{dt}+4\,m^{2}h_\alpha^2}\label{eqp2gw} 
\end{eqnarray}
respectively.\\
In terms of the Fourier amplitude $h_{\alpha,k}$ defined by
\be{fourGW}
h_{\alpha}(\vec x,t)=\int \frac{d^3k}{\pa{2\pi}^{3/2}}\;e^{i\vec k\cdot \vec x}\pa{\hat a_{\vec k}h_{\alpha,k}+\hat a_{-\vec k}^{\dagger}h_{\alpha,k}^{*}}
\ee 
one finds that (\ref{eqp1gw}) can be written as
\be{frpergwk}
\frac{\gamma}{2}\frac{m^{2}}{H^{2}}\sum_{\alpha=+,\times}\frac{1}{2\pi^{2}}\int dk\;k^{2}\pa{\ab{h_{\alpha,k}}^2}
\ee
and (\ref{eqp2gw}) as
\begin{eqnarray}
&&\frac{1}{6H^{2}}\sum_{\alpha=+,\times}\frac{1}{2\pi^{2}}\int dk\;k^{2}\paq{-\ab{\dot h_{\alpha,k}}^2\right.\nonumber\\
&&\left.+\pa{5\,\frac{k^{2}}{a^2}+4\,m^{2}}\ab{h_{\alpha,k}}^2+6H\frac{d\ab{h_{\alpha,k}}^{2}}{dt}}.\label{acpergwk}
\end{eqnarray}
The equation of motion for the tensor perturbations modes can be derived from (\ref{gwaction}) and it is given by
\be{gravdyn}
\ddot h_{\alpha,k}+3H\dot h_{\alpha,k}+\pa{\frac{k^{2}}{a^{2}}+m^{2}}h_{\alpha,k}=0.
\ee
The standard quantization procedure for the gravitational waves  is again performed by introducing the conformal time and defining $H_{\alpha,k}\equiv\sqrt{\frac{\gamma}{2}}\,a\,\sigma_{0}h_{\alpha,k}$ in terms of which one obtains the canonically normalized action for the rescaled perturbations:
\begin{eqnarray}
S_{H}&=&\frac{1}{2}\sum_{\alpha=+,\times}\int d^{3}k\,d\eta\paq{(H_{\alpha,k}')^{2}-\pa{k^{2}+a^{2}m^{2}\right.\right.\nonumber\\
&&-\left.\left.2\mathcal{H}^{2}}H_{\alpha,k}^{2}}.\label{Htenaction}
\end{eqnarray}
Such an action has the same form as (\ref{psiscaction}) for the scalar modes. The extension of the results obtained for the scalar perturbation to the tensorial case is then straightforward. Since $\sigma_{0}$ is constant at leading order, the BD and AF vacuum contributions for the tensor case are those of the scalar case multiplied by the factor $2\pa{1+6\gamma}/\pa{\gamma\,\sigma_{0}^{2}}$ owning to the different normalization factors in the definitions of $\psi_{k}$ and $H_{\alpha,k}$ in terms of $\delta\sigma_{k}$ and $h_{\alpha,k}$ respectively. We are then led to the following renormalized expressions for the BD case:
\be{m2h2BD}
\frac{1}{2\pi^{2}}\int dk\,k^{2}\pa{m^{2}\ab{h_{\alpha,k}}^{2}}\rightarrow\frac{61H^{4}}{120\,\gamma\,\sigma_{0}^{2}\pi^{2}},
\ee
\begin{eqnarray}
&&\frac{1}{2\pi^{2}}\int dk\,k^{2}\paq{\ab{\dot h_{\alpha,k}}^{2}+\pa{\frac{k^{2}}{a^{2}}+m^{2}}\ab{h_{\alpha,k}}^{2}}\nonumber\\
&&\rightarrow\frac{61H^{4}}{240\,\gamma\,\sigma_{0}^{2}\pi^{2}},\label{2rhohBD}
\end{eqnarray}
\be{hdhBD}
\frac{1}{2\pi^{2}}\int dk\,k^{2}\pa{\frac{d\,\ab{h_{\alpha,k}}^{2}}{dt}}\rightarrow0,
\ee
\be{Tk2a2h2BD}
\frac{1}{2\pi^{2}}\int dk\,k^{2}\pa{\frac{k^{2}}{a^{2}}\ab{h_{\alpha,k}^{2}}}\rightarrow-\frac{61H^{4}}{160\,\gamma\,\sigma_{0}^{2}\pi^{2}},
\ee
and to the following results for the AF vacuum
\begin{eqnarray}
&&\frac{1}{2\pi^{2}}\int dk\,k^{2}\paq{\ab{\dot h_{\alpha,k}}^{2}+\frac{k^{2}}{a^{2}}\ab{h_{\alpha,k}}^{2}}\nonumber\\
&&\rightarrow-\frac{119H^{4}}{240\,\gamma\,\sigma_{0}^{2}\pi^{2}},\label{2rhohAF}
\end{eqnarray}
\be{k2a2h2AF}
\frac{1}{2\pi^{2}}\int dk\,k^{2}\pa{\frac{k^{2}}{a^{2}}\ab{h_{\alpha,k}}^{2}}\rightarrow-\frac{119H^{4}}{160\,\gamma\,\sigma_{0}^{2}\pi^{2}}.
\ee
\section{Quantum Back-Reaction of the perturbations}
The combined back-reaction of the scalar plus tensor fluctuations determines a deviation from the first order homogeneous solution (\ref{hsol})\footnote{Let us note that small deviations from the attractor (\ref{hsol}) have been calculated to be (\ref{de1CV},\ref{ep1CV}) and are negligible for our purposes.}. Such a deviation can be expressed in terms of the slow roll parameters $\de{1}$ and $\ep{1}$ which appear on the l.h.s. of eqs. (\ref{eqp1},\ref{eqp2}) when both scalar and tensor contributions are considered. For $\ep{1}\ll1$, $\de{1}\ll 1$ (both constant) and $\de{2}$ negligible one can approximate the l.h.s. of eqs. (\ref{eqp1},\ref{eqp2}) by
\be{lhs1}
\de{1}\pa{1+6\gamma}\pa{3+2\de{1}+\de{2}-\ep{1}}\simeq3\,\de{1}\pa{1+6\gamma}
\ee
and
\be{lhs2}
2\pa{\de{1}+\ep{1}}-\de{1}\paq{\frac{4\gamma+1}{\gamma}\de{1}+2\pa{\de{2}-\ep{1}}}\simeq2\pa{\de{1}+\ep{1}}.
\ee
Summing up all the contributions on the r.h.s. one finally obtains results valid to the first order in the slow-roll approximation. In particular for BD vacuum one finds
\be{d1BD}
\de{1}=\frac{61}{12960\,\pi^{2}}\cdot\frac{\lambda\pa{1+3\,\gamma}}{\gamma\pa{1+6\,\gamma}},
\ee
\be{e1BD}
\ep{1}=-\frac{61}{1658880\,\pi^{2}}\cdot\frac{\lambda\pa{384\,\gamma^{2}+44\,\gamma-3}}{\gamma^{2}\pa{1+6\,\gamma}}
\ee
and for the AF vacuum one gets
\be{d1AF}
\de{1}=\frac{119}{51840\,\pi^{2}}\cdot\frac{\lambda}{\gamma\pa{1+6\,\gamma}},
\ee
\be{e1AF}
\ep{1}=-\frac{119}{1658880\,\pi^{2}}\cdot\frac{\lambda\pa{972\,\gamma+125}}{\gamma^{2}\pa{1+6\,\gamma}}.
\ee
Let us note that $\de{1}$ is positive in both cases indicating that the scalar field slowly increases. On the other hand $\ep{1}$ is negative for $\gamma>\gamma_{M}\equiv\pa{\sqrt{409}-11}/192\sim 0.048$ and positive elsewhere for the BD case and is negative independently of $\gamma$ in the AF case.\\
\begin{figure}[t!]
\centering
\epsfig{file=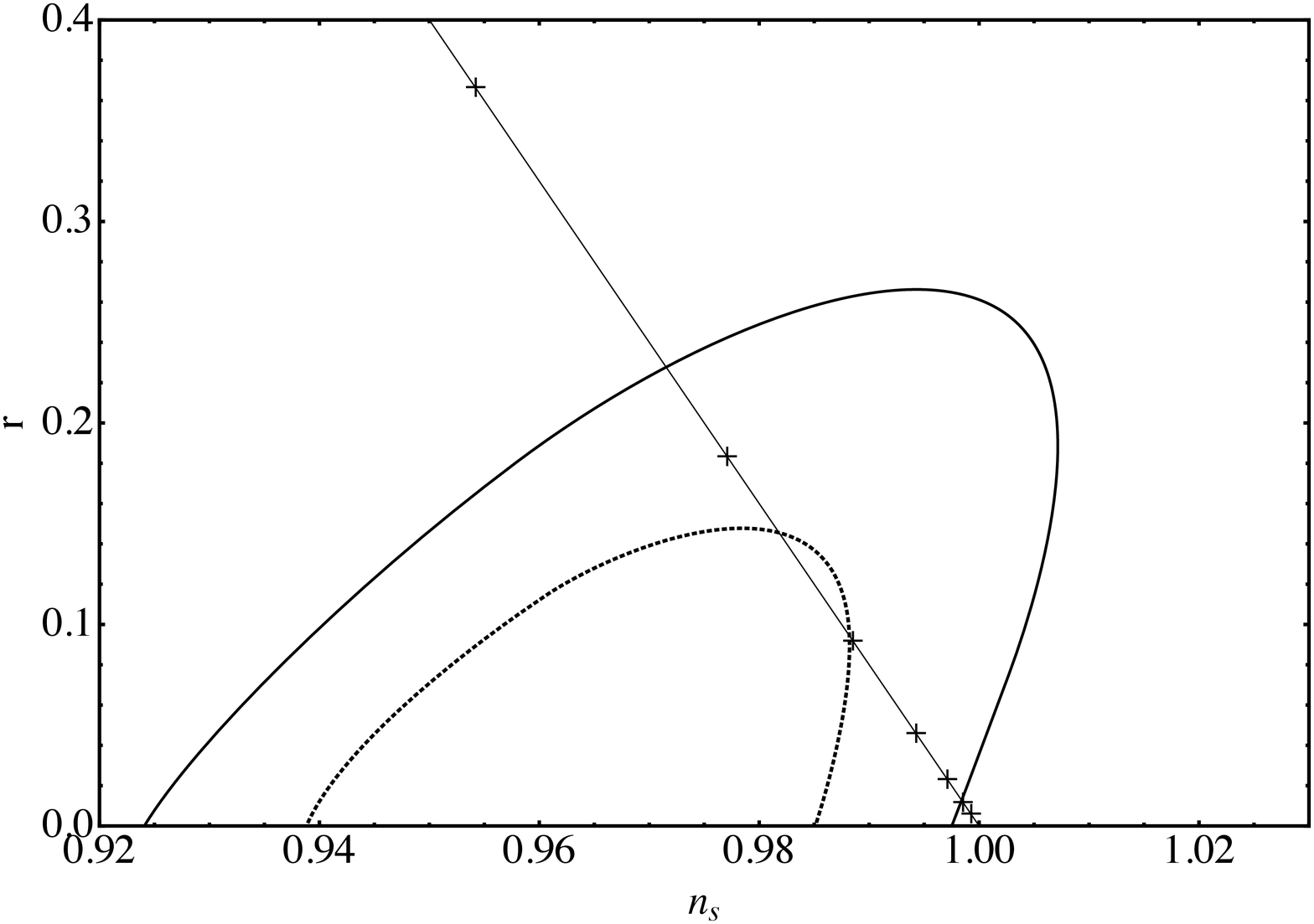, width=8.5 cm}
\epsfig{file=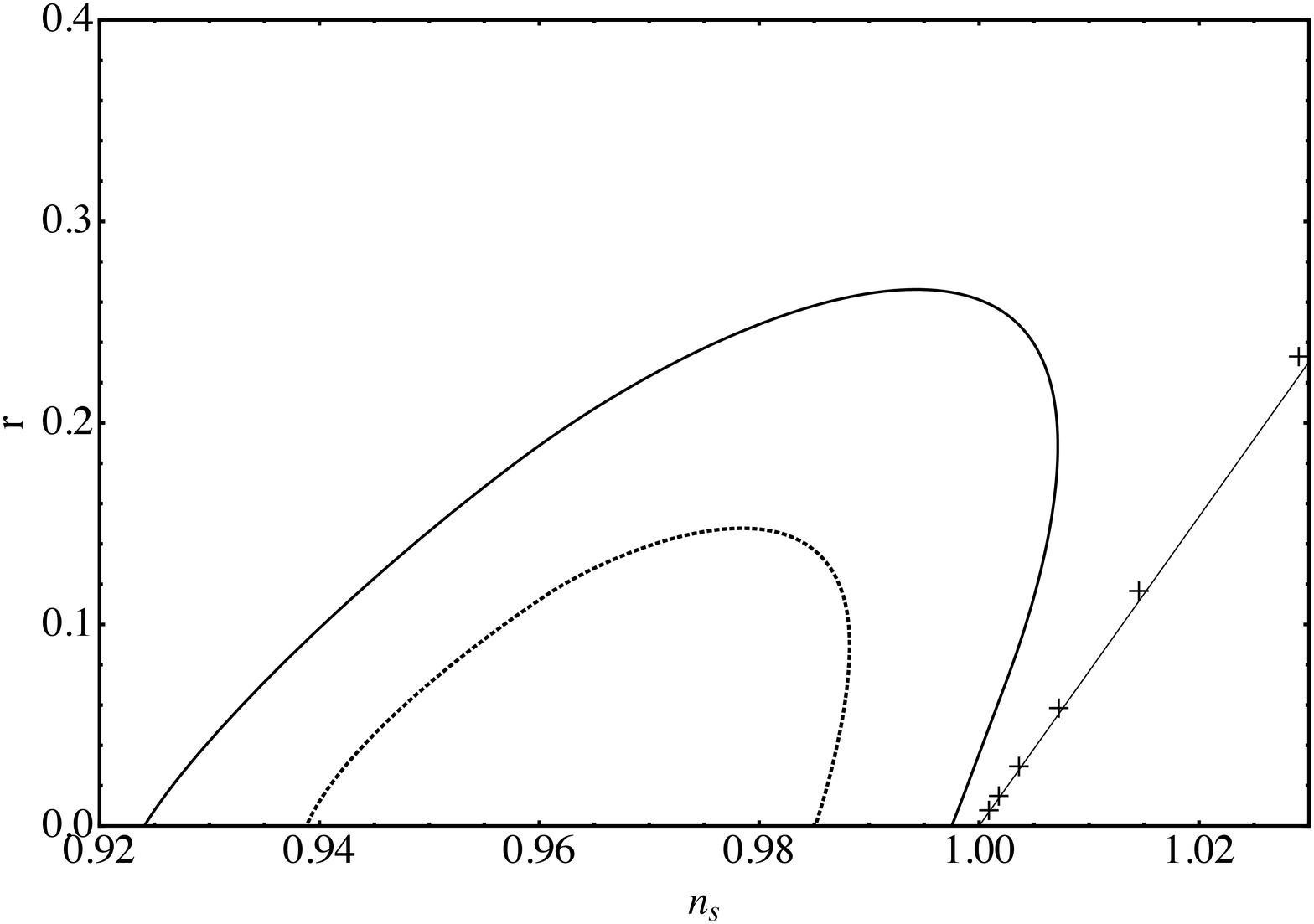, width=8.5 cm}
\caption{{\it The trajectories of the vector $(n_{s},r)$ obtained by varying $\lambda/\gamma^{2}$ for the BD vacuum (plot above) and the AF vacuum (plot below) compared with observations. The part of the trajectory between the continuous contour and the dotted contour is compatible with observations at the $68\%$ confidence level. The part of the trajectory inside the dotted contour is compatible with observations at the $95\%$ confidence level. The points on the trajectory of the above plot are for $\lambda/\gamma^{2}=2^{i}$ and $i=5,6,7,8,9,10,11$ (from the right to the left). The points on the trajectory of the plot below are for $\lambda/\gamma^{2}=2^{i}$ and $i=-1,0,1,2,3,4$ (from the left to the right).}
\label{rns}}
\end{figure}
The above SR parameters can be related to the basic observed features of the CMB spectrum and the parameters $\lambda$ and $\gamma$ constrained by observations. Values of $\gamma$ leading to a negative $\ep{1}$ are associated with an increasing Hubble parameter $H$. Such a behavior for $H$ is somehow unnatural during inflation when one generally expects that it decreases with time increasing. In IG framework \cite{Cerioni} one finds that the spectrum of the comoving curvature perturbation $\PR$ has an amplitude given by 
\be{ampPR}
\PR(k_{*})\simeq\frac{H^{2}}{4\pi^{2}\pa{1+6\gamma}\de{1}^{2}\sigma_{0}^{2}}
\ee
where all the quantities on the r.h.s. should be calculated when the pivot scale $k_{*}$ exits the horizon and for our analysis they can be treated as constant. Observational bounds coming from CMB data \cite{Komatsu:2008hk} constrain (\ref{ampPR}) to be
\be{obsPR}
\PR=\pa{2.445\pm0.096}\times 10^{-9}.
\ee
The spectral index $n_{s}-1$ of the $\PR$ spectrum is implicitly defined around the pivot scale by
\be{defns}
\PR(k)=\mathcal{P}_{\mathcal{R}}(k_*)
\left( \frac{k}{k_*} \right)^{n_s-1},
\ee
and, in IG, it can be expressed in terms of the SR parameters as
\be{nsm1}
n_{s}-1=-2\pa{\de{1}+\ep{1}}
\ee
when $\de{2}=0$. Such a spectral index with the tensor to scalar ratio
\be{r}
r\equiv\frac{\mathcal{P}_{h}}{\PR}=16|\de{1}+\ep{1}|
\ee
can be compared with CMB observations as well. In the above definition $\mathcal{P}_{h}$ is the amplitude of the spectrum of the tensor perturbations and all the quantities in (\ref{r}) should be evaluated near $k_{*}$.\\ 
At the leading order (without back-reaction) a quartic potential has a pure de Sitter attractor and (\ref{ampPR}) is singular. Such a singularity generally appears when inflation is studied in the limit $H=const$ and $\dot \sigma=0$ both in the Einstein Gravity (EG) and in the IG frameworks. This regime is often addressed as unphysical because $H$, being exactly constant, would imply a never-ending inflation and in EG is related to the presence of a cosmological constant which does not produce any curvature perturbation. In the IG framework, however, the $H=const$ regime is the consequence of the scalar field dynamics and in this context scalar fluctuations are indeed generated. The homogeneous dynamics can then be modified by the quantum corrections and if such corrections become large enough or when other effects, such as the interaction with other fields, become important inflation may come to an end. 
In our perturbative approach we are not able to describe these effects beyond the linear order but we may safely assume that the quantum fluctuations determine the slow-roll parameters before the accelerated era ends when they are no longer small.\\ 
The singular behavior of $\PR$ is closely related to the definition of the gauge invariant variable $\mathcal{R}$ when $\dot \sigma=0$. This means that the amplitude of the scalar perturbations generated during inflation needs a different gauge treatment to be properly studied in this dynamical regime. Exploring the consequences of a choice of a different variable for the description of the evolution of the scalar perturbations is not so straightforward and goes beyond the scope of this paper. We just emphasize that our perturbative approach is not suitable for calculating the quantum corrections to $\PR$ which is divergent on the attractor (\ref{hsol}) but is inverse proportional to $\de{1}$ away from it and we restrict our analysis to $n_{s}$ and $r$.\\
On imposing the observational bounds on $n_{s}$ and $r$ one can determine the values of $\gamma$ and $\lambda$ compatible with CMB data. Let us note that there exists a linear relation between (\ref{nsm1}) and (\ref{r}) of the form
\be{relnsr}
r=8|n_{s}-1|.
\ee
In particular, for the BD vacuum, independently of $\lambda$, when $0<\gamma<\gamma_{M}$, one has $n_{s}<1$  and (\ref{relnsr}) can be re-written as $r=8\pa{1-n_{s}}$ and when $\gamma>\gamma_{M}$ one finds $n_{s}>1$  and (\ref{relnsr}) becomes $r=8\pa{n_{s}-1}$. The latter relation always holds for the AF vacuum choice. The consistency relation (\ref{relnsr}) is plotted in the Figure (\ref{rns}) and compared with observations. Such a comparison shows that only the BD vacuum corrections with $\gamma<\gamma_{M}$ are compatible with the observed spectrum of the CMB anisotropies.\\
In particular, when $\gamma\ll1$, one finds
\begin{eqnarray}
&&n_{s}^{(BD)}-1\simeq-\frac{61}{276480\pi^{2}}\frac{\lambda}{\gamma^{2}},\\ 
&&n_{s}^{(AF)}-1\simeq\frac{2975}{165888\pi^{2}}\frac{\lambda}{\gamma^{2}}
\end{eqnarray}
and
\begin{eqnarray}
&&r^{(BD)}\simeq-\frac{61}{34560\pi^{2}}\frac{\lambda}{\gamma^{2}},\\ 
&&r^{(AF)}\simeq\frac{2975}{20736\pi^{2}}\frac{\lambda}{\gamma^{2}}.
\end{eqnarray}
which simply depend on $\lambda/\gamma^{2}$. Different choices of the ratio $\lambda/\gamma^{2}$ are then presented in the Figure (\ref{rns}).\\
The BD vacuum turns to be compatible with CMB data for $\gamma\ll1$ and $\lambda$ such as 
\be{consgsmall}
2^{6}\lesssim\lambda/\gamma^{2}\lesssim 2^{10}. 
\ee
For these values of the parameters the condition for inflation to occur, $\ep{1}\simeq 10^{-5} \lambda/\gamma^{2}<1$, is always satisfied and one also has $\ep{1}>0$ (no super-acceleration). On the other hand AF vacuum results are not compatible with $(n_{s},r)$ observational constraints (see Fig. (\ref{rns})).
Even for tiny values of $\lambda$ the effect of the quantum back-reaction is non negligible providing $\gamma$ is small enough (let us remember that $\gamma\rightarrow 0$ corresponds to the limit of Einstein gravity). On taking into account quantum effects through the adoption of a flat-space inspired Coleman-Weinberg potential \cite{CFTV,Cerioni} one finds $\lambda/\gamma^{2}\simeq 10^{-10}/\gamma$ for $\gamma\ll 1$ which, depending on $\gamma$, may still be large.\\ 
\begin{figure}[t!]
\centering
\epsfig{file=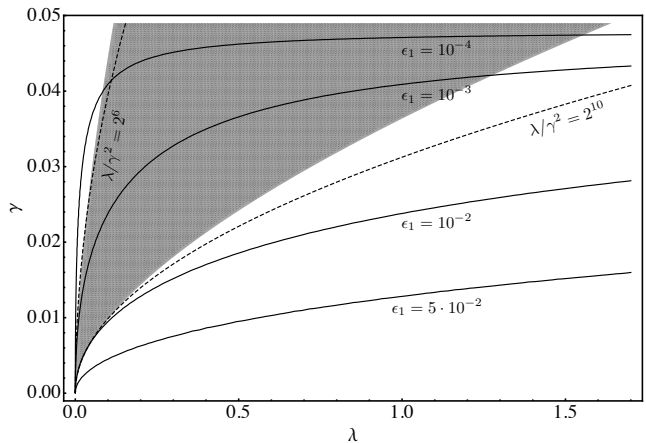, width=8.5 cm}
\caption{{\it The shaded region represents the observational bounds on $(\lambda,\gamma)$ for the BD vacuum. The solid lines are some of the curves for $\ep{1}=const$ and the dashed lines are the contours of the region (\ref{consgsmall}) found for $\gamma\ll 1$. These contours deviate more and more from the boundaries of the shaded region as $\gamma$ increases.}
\label{GL}}
\end{figure}
The more general case ($\gamma<\gamma_{M}$) for the BD vacuum is plotted in Fig. (\ref{GL}) where we impose the observational bounds on $(n_{s},r)$ which appear in Fig. (\ref{rns}) to $(\lambda,\gamma)$. Such bounds are obtained by calculating the intersection between the line (\ref{relnsr}) in the $(n_{s},r)$ plane and the $68\%$ confidence level contour. They can be then expressed as constraint on $n_{s}$ of the form $0.972\lesssim n_{s}\lesssim 0.998$ in the $(\lambda,\gamma)$ plane. Figure (\ref{GL}) shows the possible choices of $(\lambda,\gamma)$ compatible with observations compared with the corresponding value of $\ep{1}$. We find that only values of $\ep{1}$ such as $0<\ep{1}<10^{-2}$ are compatible with observations. Moreover the dashed lines in Fig. (\ref{GL}) represent the constraints (\ref{consgsmall}) on $\lambda/\gamma^{2}$ found analytically for $\gamma\ll1$.\\
To conclude, deviations from scale invariance compatible with observations can be found for the BD vacuum, $\gamma<\gamma_{M}$ and $\lambda$ chosen according to the constraints plotted in Fig. (\ref{GL}).\\
\section{Conclusions}
The back-reaction of the quantum scalar and tensor perturbations on the homogeneous dynamics of an IG inflationary model with a quartic, self-interacting potential have been studied. The calculations were performed perturbatively by expanding around the de Sitter attractor of the homogeneous dynamics and the energy-momentum tensor of the quantum fluctuations was renormalized through the dimensional regularization and fourth order adiabatic subtraction both for the BD and the AF vacuum. We found that the back-reaction for the BD vacuum choice was able to reproduce a nearly scale invariant spectrum for the scalar perturbations compatible with observations for suitable choices of the free parameters of the theory $\gamma$ and $\lambda$. On the other hand the AF vacuum was not compatible with the observed spectrum of the CMB.\\ 
We emphasize that our radiative corrections during inflation have been evaluated in a de Sitter space (in contrast with previous Coleman-Weinberg flat-space results \cite{CFTV,Cerioni}) and that we have shown that such quantum corrections can lead to non scale invariant inflationary spectra compatible with observations.
\acknowledgments
We would like to thank A. Cerioni, F. Finelli, G. Marozzi and G.P. Vacca for useful comments and suggestions.

\end{document}